\documentclass[useAMS,usenatbib]{mnras}

\usepackage{amsmath}
\usepackage{amsfonts}
\usepackage{amssymb}
\usepackage{graphicx}
\graphicspath{{Figures/}}

\usepackage[normalem]{ulem}
\usepackage{RefJ}

\usepackage{enumitem}

\usepackage{hyperref}
\usepackage[]{natbib}

\newcommand{\dd}{\mathrm{d}}

\DeclareMathOperator{\tr}{tr}

\usepackage{booktabs} 

\let\vec\boldsymbol 

\usepackage{color}

\usepackage{array}
\newcolumntype{T}{>{\tiny}l} 
\newcolumntype{H}{>{\Huge}l} 

\title[An intuitive parametric model for 3D compressible (M)HD turbulence]{An intuitive parametric model for 3D compressible hydrodynamical and MHD turbulence}
\author[J.-B. Durrive, K. Ferri\`ere, and P. Lesaffre]{Jean-Baptiste Durrive$^{1,2,3}$\thanks{E-mail:jdurrive@protonmail.com}, Katia Ferri\`ere$^{1}$, and Pierre Lesaffre$^{2}$\\
$^{1}$ Institut de recherche en astrophysique et plan\'etologie - Universit\'e Toulouse III - Paul Sabatier, Observatoire Midi-Pyr\'en\'ees,\\
Centre National de la Recherche Scientifique, UMR5277 - France\\
$^2$ Laboratoire de physique de l'\'ENS, \'Ecole normale sup\'erieure de Paris, Centre National de la Recherche Scientifique, FR684,\\
Universit\'e Paris Diderot-Paris 7, Sorbonne Universit\'e, UMR8023 - France\\
$^3$ Centre for mathematical Plasma-Astrophysics, Celestijnenlaan 200B, 3001 Leuven, KU Leuven, Belgium\\}

\begin{document}

\date{Accepted --- ; Received --- ; in original form ---}

\pagerange{\pageref{firstpage}--\pageref{lastpage}} \pubyear{2020}

\maketitle

\label{firstpage}

\begin{abstract}
An analytical model for three-dimensional incompressible turbulence was recently introduced in the hydrodynamics community which, with only a few parameters, shares many properties of experimental and numerical turbulence, notably intermittency (non-Gaussianity), the energy cascade (skewness), and vorticity alignment properties.
In view of modeling astrophysical environments, we introduce a manner to extend to compressible fluids the three-dimensional turbulent velocity field model of \cite{ChevillardEtAl10}, as well as the three 3D turbulent magnetic field models of \cite{DurriveEtAl20}, following the same procedure based on the concept of multiplicative chaos.
Our model provides a complementary tool to numerical simulations, as it enables us to generate very quickly fairly realistic velocity fields and magnetic fields, the statistics of which are controllable with intuitive parameters. Therefore our model will also provide a useful tool for observers in astrophysics.
Finally, maybe even more than the model itself, it is the very procedure that matters the most: our model is modular, in the sense that it is constructed gradually, with intuitive and physically motivated steps, so that it is prone to many further improvements.
\end{abstract}


\section{Introduction}

Fluids in nature, magnetized or not, are most often in a turbulent state, up to the largest cosmological scales and earliest epochs of the Universe \citep[e.g.][]{Subramanian19}. Hence, in order to describe natural phenomena, from laboratory to astrophysical environments, it is crucial to be able to model turbulence. But the difficulty to do so precisely is well-known: the solutions of the Navier-Stokes equations and the equations of magnetohydrodynamics (MHD) are still far from reach analytically, and generating realistic data by means of numerical simulations is very costly, both in terms of time and of resources.

In the quest for finding explicit solutions to the incompressible Navier-Stokes equations, \cite{ChevillardEtAl10,ChevillardEtAl11,ChevillardEtAl12,ChevillardEtAl13,ChevillardHDR,PereiraEtAl16,PereiraEtAl18} built a powerful (i.e. concise and yet efficient) model for incompressible hydrodynamical turbulence, based on the concept of Gaussian multiplicative chaos. This concept was first introduced by the mathematician J.-P. Kahane in \cite{Kahane85}. It is closely related to fractals \citep[e.g.][]{Mandelbrot72}, which is why it is a natural tool to model processes comprising self-similarity, such as turbulence. It is suited for many other applications, but our focus here is (magneto)fluid dynamics in view of modeling astrophysical environments. For a recent review on Gaussian multiplicative chaos, see for example \cite{RhodesVargas14}.
Mathematically speaking, Gaussian multiplicative chaos consists in considering the exponential of a Gaussian field with logarithmic covariance. Simply said, what lies at the heart of our model is the fact that, while the sum (or more generally a linear operation) of Gaussian processes is a Gaussian process, the product of Gaussian processes is non-Gaussian, i.e. intermittent. Hence the terminology `multiplicative' in Gaussian multiplicative chaos, and the choice of the non-linear operation that the exponential is.

Based on the aforementioned series of papers, we proposed in \cite{DurriveEtAl20} three turbulence models for the magnetized case, i.e. for three dimensional MHD turbulence. As in \cite{ChevillardEtAl10}, we limited ourselves to the incompressible case. In the present paper, we propose generalizations of this method to include compressibility, for both unmagnetized and magnetized fluids.

Our objective follows that of \cite{DurriveEtAl20}: we try to adapt tools developed in the hydrodynamics and MHD communities to the needs of the astrophysics community. In that respect, our priority while designing the present model, has been to try and make it as practical as possible.

Firstly, given its construction, this model should, hopefully, be useful for both numerically-oriented and observationnally-oriented researchers. Indeed, on the one hand, one can use this model to initialize numerical simulations, with quickly generated non-trivial fields. It may also serve as prescriptions for fields at unresolved scales. On the other hand, thanks to the free parameters controlling the properties of the fields, our model may be used to characterize statistically data (stemming from either observations or simulations) or to assess statistical error bars by generating hundreds of realizations.
We are currently implementing numerically this model, and in a close future we will make a public release of our code. A crucial feature of this approach is that it is very efficient numerically, because most of the formulae are nothing but convolutions, i.e. simple products in Fourier space. We may generate fields in a few minutes, which, with the same resources and at the same resolution, are obtained in typically several days or weeks by direct numerical simulations (DNS). Obviously DNS results remain more realistic in the details, but with our approach the statistics of the fields are easily controllable. In that sense our approach is complementary to DNS.

Secondly, we made an effort to present the model pedagogically. Hence, while a large part of the derivation is based on mathematical considerations, readers in a hurry or those only looking for a practical tool, do not need to delve into any of these technicalities. Each technical argument is accompanied with a physical meaning, so that the formulae ultimately are very intuitive. Naturally we also added numerous references to the literature, to satisfy as well readers who, on the contrary, are interested in the fundamental aspects of the model. Indeed, practicalities aside, this approach is obviously also interesting per se, from a theoretical point of view, as it takes part in the studies on how the concept of fractals and multiplicative chaos may help modeling turbulence.

Our notations will follow closely those used in \cite{DurriveEtAl20}: our work is a sequel of it, and we want to keep the description focused on physical arguments. This is done at the cost of losing some mathematical rigor. The reader caring about the technicalities hidden by these more intuitive notations is invited to have a look at the first appendix of \cite{DurriveEtAl20}. Now, in order to make the present paper as self-consistent as possible, let us recall that (i) we use everywhere the shorthand notation
\begin{equation}
\vec{r} \equiv \vec{x}-\vec{y} \hspace{0.5cm} \text{and} \hspace{0.5cm} r \equiv |\vec{r}|,
\label{shortHandNotations}
\end{equation}
as most expressions are convolutions, with $r$ corresponding to the distance from a given position $\vec{x}$ when integrating with respect to $\vec{y}$, (ii) we use subscripts $d, v, \omega$ and $m$ for quantities associated with density fields, velocity fields, vorticity fields and magnetic fields respectively, and (iii) variables with a tilde denote random variables and subscript $g$ is added to random fields with Gaussian statistics.

The paper is organized as follows. First, in section~\ref{sec:Nutshell}, we give a brief summary of the paper, to help readers keep track of the logic behind the various steps of the derivation once they delve into the detailed calculations. Then, in section~\ref{sec:Dynamics}, we present the equations governing the dynamics of the fields of which we want to build a stochastic representation. Section~\ref{sec:Hydro} is dedicated to the model in the hydrodynamical case, and section~\ref{sec:MHD} to the magnetized case. We conclude and give a glimpse of our prospects in section~\ref{sec:Conclusion}.

\section{Outline in simple words}
\label{sec:Nutshell}

To improve the presentation, let us summarize, using words only, the adopted method.

Given the observed stochastic nature of turbulence, stochastic calculus\footnote{However, let us straightaway reassure any concerned reader that no prior knowledge of this specialized mathematical topic is needed to understand our paper, as detailed in the first appendix of \cite{DurriveEtAl20}.} is a most natural formalism to describe turbulent fields.
Starting from the equations of motion, we derive formulae telling us how fluid particles are being deformed over time. We then transform these (approximate) solutions into random fields by setting their initial conditions as Gaussian random fields\footnote{More precisely, this randomization is only one step out of a five-step procedure that we call `turbulization'. The other steps introduce an injection scale, a dissipation scale, a parameter quantifying how `rough' the turbulence is, and a parameter quantifying the intermittency. For details, see section \ref{sec:turbulization}.}.
Randomness is thus implemented through the initial conditions, while non-Gaussian features emerge from the dynamical evolution.

More specifically, the model of \cite{ChevillardEtAl10} for an incompressible velocity field is derived as follows. Due to incompressibility, Biot-Savart's law provides a one-to-one correspondence between velocity and vorticity, similar to the one-to-one correspondence between magnetic field (which is divergence-free) and current density. Using this fact, in this model what is constructed (`by hand') is the vorticity field\footnote{Doing so is natural since vortex tubes have long been identified as a key quantity in modeling hydrodynamical turbulence \citep{SheEtAl90,FargeEtAl01,ApolinarioEtAl20}.}, and the velocity field is then deduced by means of Biot-Savart's law.
Accordingly the authors start by studying the dynamics of vorticity. They notice that on a short enough timescale $\tau$ (namely the correlation timescale of velocity gradients), an efficient approximate solution of the evolution equation for vorticity can be found. Thus they derive an explicit formula for how vortices are being stretched in the fluid.
To complete this formula, the initial state needs to be specified: they choose it to be a Gaussian random field, in order to take into account phenomenologically the resultant of all possible fluctuations induced locally by the many surrounding eddies, by a large-scale eddy and/or by some external forcing.
Biot-Savart's law then translates this vorticity model into a model for the velocity field. Hence, the velocity field is modeled as the result of the random stretching (dictated by the equation of motion) of an initially Gaussian vorticity field.

Then, the velocity field thus constructed is considered as a random field representative of the stationary state of the fluid. Indeed, $\tau$ being a correlation timescale, at each timestep $\tau$ the fluid gets deformed and then decorrelates, over and over again, such that the stationary state of the fluid ends up having similar statistics as this velocity field deformed during the time lapse $\tau$.
In addition, since the parameter $\tau$ corresponds to the time lapse during which stretching occurs, $\tau$ controls how much the velocity field deviates from its initial state. The latter being Gaussian, the parameter $\tau$ quantifies the intermittency of the velocity field (intermittency being defined as departure from Gaussianity).
By simply tuning this intermittency parameter $\tau$, this velocity field model shares many properties of experimental and numerical turbulence \citep[cf.][and the following works analyzing these fields]{ChevillardEtAl10}.

The models of \cite{DurriveEtAl20} for incompressible MHD turbulence are derived in a similar manner, but starting from the MHD equations. The main new ingredient with magnetic fields is that flux tubes are not only stretched and contracted as vortices are, but they are also being sheared by vorticity.

In the above incompressible cases, the velocity field is fully characterized by its vorticity field, which, by construction, is divergence-free and thus contains only two degrees of freedom. Therefore, in the description, one needs to fix only two degrees of freedom.
Indeed, while the deformation itself is obtained from first principles (it is dictated by the equations of the dynamics), the fact that vorticity is modeled as a Gaussian field that is being deformed is a phenomenological input, which fixes these degrees of freedom.
Now, in the present work we extend this approach to take compressibility into account. In this case, Biot-Savart's law, generalized to a compressible fluid, expresses the velocity field not only in terms of the vorticity field, but also in terms of the dilatation field. Hence, a third degree of freedom now comes into play, and an additional phenomenological input is required. Guided by astrophysical data, in which density fields tend to be lognormal \citep[e.g.][]{LevrierEtAl18}, we suggest modeling the dilatation field as a fractional Gaussian field (i.e. a Gaussian field of which the statistics are controllable with a few parameters). Having fixed this third degree of freedom, we derive (astrophysically) relevant expressions for compressible hydrodynamical turbulence, as well as MHD turbulence, from a generalized Biot-Savart law.

Finally, it is noteworthy that all the random fields that we build in this model (dilatation, density, velocity, vorticity, deformation, current density, and magnetic fields) are correlated, by construction, because we manage to build them from a single Gaussian white noise vector (the Gaussian `seed' $\widetilde{\vec{s}}_g$ below).

\section{Dynamical evolution}
\label{sec:Dynamics}

We aim at constructing an analytic stochastic representation of a density field, a velocity field, a vorticity field and a magnetic field coupled together, as realistically as possible. Mathematically speaking, we are thus aiming at expliciting (approximate) solutions to the system of equations consisting of the continuity equation, momentum equation and induction equation, including viscous and resistive effects. However, a basic feature of the approach of \cite{ChevillardEtAl10} and \cite{DurriveEtAl20} is to start with the aforementioned equations in the limit of vanishing viscosity and resistivity. Indeed, such effects are later taken into account phenomenologically through various regularizations (cf. item (iii) of the turbulization procedure detailed in section \ref{sec:turbulization}). Therefore, the equations for the dynamics that we are now going to consider for constructing our formulae are the following.

\subsection{Density field}

Denoting the material derivative (i.e. evaluated while moving with
the fluid) as
\begin{equation}
\frac{\dd}{\dd t} \equiv \partial_t + \vec{v} \cdot \vec{\nabla},
\end{equation}
the conservation of mass of a fluid of density $\rho$ and velocity $\vec{v}$ can be described by
\begin{equation}
\frac{\dd \rho}{\dd t} = - \vartheta \rho,
\label{MassConservation}
\end{equation}
where $\vartheta$ is the dilatation field,
\begin{equation}
\vartheta \equiv \vec{\nabla} \cdot \vec{v}.
\label{def:dilatation}
\end{equation}
This field $\vartheta$ expresses local expansion or contraction, according to whether it is positive or negative respectively. It is equal to zero in the incompressible case of \cite{DurriveEtAl20}, and is the new key ingredient of the present compressible case. 

\subsection{Velocity field}

The momentum evolution equation for an inviscid fluid of pressure $p$ in a magnetic field $\vec{B}$ and gravitational field $\vec{g}$ reads
\begin{equation}
\frac{\dd \vec{v}}{\dd t} = -\frac{1}{\rho} \vec{\nabla} p + \frac{1}{\rho} \vec{j} \times \vec{B} +\vec{g}.
\label{EulerEquation}
\end{equation}
The terms in the right-hand side correspond to the forces acting on the fluid element, namely the pressure gradient force, the Lorentz force, and the gravitational force respectively. The current density $\vec{j}$ is given by Amp\`ere's law,
\begin{equation}
\vec{j} = \frac{1}{\mu_0} \vec{\nabla} \times \vec{B},
\end{equation}
where $\mu_0$ is the permeability of free space.

Equation \eqref{EulerEquation} identifies the physical origin of motion, falling into the category of dynamics. It is at the heart of our model, but in fact a large part of our formalism deals with the fluid deformation and motion, without concern about the causes of these deformation and motion, falling into the category of kinematics.
Indeed, our starting point for modeling the velocity field is the Helmholtz decomposition,
\begin{equation}
\vec{v}=\vec{v}_c+\vec{v}_r,
\label{def:v}
\end{equation}
where the first term, called the compressional component, satisfies $\vec{\nabla} \times \vec{v}_c = \vec{0}$ and is related to the dilatation field \eqref{def:dilatation} through
\begin{equation}
\vec{v}_c(\vec{x}) = \frac{1}{4 \pi} \int_{\mathbb{R}^3} \frac{\vartheta \vec{r}}{r^3} \dd V,
\label{def:vc}
\end{equation}
while the second term, called the rotational component, satisfies $\vec{\nabla} \cdot \vec{v}_r = 0$ and can be written as
\begin{equation}
\vec{v}_r(\vec{x}) = \frac{1}{4 \pi} \int_{\mathbb{R}^3} \frac{\vec{\omega} \times \vec{r}}{r^3} \dd V,
\label{def:vr}
\end{equation}
(fluid equivalent of the magnetic Biot-Savart law) where the vorticity field is defined as
\begin{equation}
\vec{\omega} \equiv \vec{\nabla} \times \vec{v}.
\end{equation}
In these expressions we do not consider boundary terms, assuming the fields vanish sufficiently fast at infinity, i.e. far away from the region of interest.

Readers more familiar with electrostatics and magnetostatics than with hydrodynamics will get a feeling for where these expressions come from thanks to the following comparison. The compressional component $\vec{v}_c$ is formally equivalent to the electric field, $\vec{E}$, which satisfies Maxwell's equations $\vec{\nabla} \times \vec{E} = \vec{0}$ and $\vec{\nabla} \cdot \vec{E} = \rho_e/\epsilon_0$ (with $\rho_e$ the electric charge density and $\epsilon_0$ the permittivity of free space), and which can be written as
\begin{equation}
\vec{E} = \frac{1}{4 \pi} \int \frac{(\rho_e/\epsilon_0) \ \! \vec{r}}{r^3} \dd V,
\end{equation}
whereas $\vec{v}_r$ is analogous to the magnetic field, $\vec{B}$, which satisfies $\vec{\nabla} \cdot \vec{B} = 0$ and $\vec{\nabla} \times \vec{B} = \mu_0 \vec{j}$, and which can be written as the original (magnetic) Biot-Savart law
\begin{equation}
\vec{B} = \frac{1}{4 \pi} \int \frac{(\mu_0 \vec{j}) \times \vec{r}}{r^3} \dd V.
\label{BiotSavart_magneticField}
\end{equation}

\subsection{Gradient of the velocity field}

A central quantity of our model is the gradient of the velocity field
\begin{equation}
\vec{G} \equiv \left(\vec{\nabla} \vec{v}\right)^\textsc{T},
\end{equation}
where $\textsc{T}$ denotes transposition. Taking the gradient of the Helmholtz decomposition \eqref{def:v}, one can write $\vec{G}$ as \citep{WuMaZhou07}
\begin{equation}
\vec{G} = \vec{\mathcal{D}} + \vec{\Omega},
\label{def:G}
\end{equation}
where the rotation tensor (i.e. the antisymmetric part of $\vec{G}$) is a cross product with vorticity, namely
\begin{equation}
\vec{\Omega} = \frac{1}{2} \vec{\omega} \times,
\label{def:Omega}
\end{equation}
and the strain-rate tensor (i.e. the symmetric part of $\vec{G}$) may itself be decomposed as
\begin{equation}
\vec{\mathcal{D}} = \vec{\mathcal{D}}_c+\vec{\mathcal{D}}_r,
\label{def:D}
\end{equation}
where the compressional component reads
\begin{equation}
\vec{\mathcal{D}}_c = \frac{1}{4 \pi} \hspace{-0.1cm} \int_{\mathbb{R}^3} \frac{(r^2 \vec{I}-3 \vec{r} \vec{r}) \vartheta }{r^5} \dd V,
\label{def:Dc}
\end{equation}
and the rotational component
\begin{equation}
\vec{\mathcal{D}}_r = -\frac{3}{8 \pi} \hspace{-0.1cm} \int_{\mathbb{R}^3} \frac{\vec{r} (\vec{\omega} \times \vec{r}) + (\vec{\omega} \times \vec{r}) \vec{r}}{r^5} \dd V.
\label{def:Dr}
\end{equation}
In these expressions products of vectors of the form $\vec{x} \vec{y}$ are tensor products ($\vec{x}\vec{y}|_{ij}=x_i y_j$). As in \cite{WuMaZhou07}, we do not include a Cauchy principal value in the expressions of \eqref{def:Dc} and \eqref{def:Dr} \citep[while it is mentioned in ][]{DurriveEtAl20} because in the following this technicality is avoided thanks to the regularizations of the kernels. One should not be misled however by the fact that as such \eqref{def:Dc} seems to yield a traceless $\vec{\mathcal{D}}_c$ while its trace should equal $\vartheta$.

In contrast to $\vec{\mathcal{D}}$, which has both a compressional and a rotational component (cf. Eq~\eqref{def:D}), $\vec{\Omega}$ (cf. Eq~\eqref{def:Omega}) has only a rotational component; its compressional component, $\vec{\Omega}_c = 1/2 \vec{\omega}_c \times$, vanishes trivially because $\vec{\omega} \equiv \vec{\nabla} \times \vec{v}_c = \vec{0}$ (cf. above Eq~\eqref{def:vc}). In other words, compressibility affects only the strain-rate tensor, $\vec{\mathcal{D}}$, not the rotation tensor, $\vec{\Omega}$.

Physically, the matrix $\vec{\mathcal{D}}$ includes vortex stretching, which is an essential kinematic mechanism. This nonlinear effect is responsible for the cascade process in turbulence, by which large-scale eddies break apart into smaller and smaller eddies which rotate faster and faster, i.e. with increasingly stronger enstrophy \citep{WuMaZhou07}.
Stretching is most efficient when $\vec{\omega}$ is aligned with the stretching principal axes of $\vec{\mathcal{D}}$, at a rate given by the eigenvalues of $\vec{\mathcal{D}}$. Therefore this process is also related to vorticity alignments in turbulent flows.
The strain rate that causes stretching can be either a background field induced by other vortices (hence the fact that the expression for $\vec{\mathcal{D}}$ involves a spatial integral), or induced locally by the vortex itself. For a study of the local contribution (induced by a vortical structure in its neighboring vicinity) versus the nonlocal contribution (background strain rate induced in the vicinity of the structure by all the remaining vorticity) in the strain rate tensor, and its implications for vorticity alignment, see for example \cite{HamlingtonEtAl08}.

The above considerations are kinematic. The dynamics of the velocity gradient, following fluid particles, is obtained by taking the gradient of the equation of motion \eqref{EulerEquation}, yielding
\begin{equation}
\frac{\dd \vec{G}}{\dd t} = -\vec{G}^2 + \vec{\nabla} \vec{\Pi},
\label{EvolutionOfG}
\end{equation}
where
\begin{equation}
\vec{\Pi} = -\frac{1}{\rho} \vec{\nabla} p + \frac{1}{\rho} \vec{j} \times \vec{B} +\vec{g}.
\end{equation}
The non-linear term $\vec{G}^2$ in \eqref{EvolutionOfG} induces gradient steepening and physically corresponds to self-stretching.
This equation is fundamental to analyze turbulent flows. Indeed, the full velocity gradient tensor $\vec{G}$ characterizes variations of all velocity components, in all directions, which constitutes valuable information, complementary to the usual studies of the statistics and anomalous scaling of longitudinal and transverse velocity increments. A vast literature exibiting its properties is available, especially in the incompressible case \citep[e.g.][]{ZeffEtAl03,ChevillardMeneveau06,Wallace09,Meneveau11} but also compressible \citep{SumanGirimaji11,ParasharEtAl17}.

Equation \eqref{EvolutionOfG} is not closed in terms of $\vec{G}$ because one still needs to prescribe $\vec{\Pi}$. This missing additional constraint on $\vec{\Pi}$ is commonly referred to as a closure relation, because it enables us to end up with a closed system of equations. The choice of closure relation is a crucial task, as it impacts the whole model and determines its domain of validity, but it is also a very challenging one. In appendix \ref{Appendix:Closures} we give some examples of closure relations that can be found in the literature specialized in turbulence, which are the relations we took inspiration from for the present work. We postpone to section \ref{sec:turbulization} the presentation and justification of the choice of closure for our model, because we need to introduce a few more elements beforehand.

\subsection{Dilatation field}
\label{sec:Dilatation}

Taking the divergence of the equation of motion \eqref{EulerEquation}, or equivalently taking the trace of \eqref{EvolutionOfG}, yields the evolution equation for the dilatation field,
\begin{equation}
\frac{\dd \vartheta}{\dd t} = - \tr(\vec{G}^2) - \vec{\nabla} \cdot \left(\frac{\vec{\nabla} p}{\rho}\right) + \vec{\nabla} \cdot \left(\frac{\vec{j} \times \vec{B}}{\rho}\right)+ \vec{\nabla} \cdot \vec{g}.
\label{Equation_dilatation}
\end{equation}
In order to better visualize the assumptions underlying our modeling, we write the above equation in full generality, but in fact in this work we will retain only the first term in the right-hand side, i.e. the one stemming from self-stretching. Let us now expose the reasons why we discard the other terms (for simplicity obviously, but not only), and let us discuss to what extent, a priori, this choice may restrict the domain of validity of our model.

As far as the pressure term is concerned, writing \eqref{Equation_dilatation} in dimensionless form, it appears that it is inversely proportional to the Mach number squared. Therefore we place ourselves in the high Mach number regime.

Regarding the Lorentz force term, we discard it as a first step, planning to increase the complexity of our model gradually and leaving the task of considering magnetic feedback for future works. That being said, we have more than a simple strategic reason. Firstly it has been observed in hydrodynamics \citep[e.g.][]{PelzEtAl85} that, at least in regions of low dissipation, the velocity and vorticity vectors tend to get aligned, statistically speaking (i.e. the probability distribution function of the cosine of the angle between the two fields peaks at $\pm 1$, while uncorrelated Gaussian variables would produce a flat distribution). This phenomenon is called Beltrami correlation or Beltramization, in reference to Beltrami vector fields which are parallel to their own curl. Now, similar correlations emerge in MHD turbulence where the fluid has a tendency to produce various correlations between the fields. Among them are Alfv\'enic correlation, where velocity and magnetic field have a tendency to align, and force-free correlation, where the magnetic field and the electric current density tend to align \citep{ServidioEtAl08,MieschEtAl15,SetaEtAl20}. Hence, the diminishing of nonlinearity in the inertial range by alignment seems common enough to legitimize the fact that we are going to build our model neglecting the feedback from the Lorentz force in \eqref{Equation_dilatation}.
Secondly, in the following we will introduce some phenomenological parameters into the dynamical equations of the present section, with the aim of widening their domain of validity. In other words, we will modify our equations a little, or `post-process' them so to speak, following a procedure detailed in section \ref{sec:turbulization}. As a consequence, while we neglect the effect of the Lorentz force in the construction at this stage, ultimately our model may still describe to some extent non force-free MHD. This procedure already proved fruitful in the hydrodynamical case \citep{ChevillardEtAl10}, in which the model is first constructed based on the inviscid limit, but in a second step regularization parameters are introduced, which ultimately account for dissipation.

Finally, regarding the gravitational acceleration term, it all depends on the environment one wants to model, because this term as such is not closed. For instance, if we were to model a self-gravitating medium, the relevant closure relation to join to our system of equations would be Poisson's equation $\vec{\nabla} \cdot \vec{g}=-4 \pi \rho$. However, we limit the scope of this paper to turbulent interstellar media such as molecular clouds in which self-gravity is subdominant, i.e. not prone to gravitational instability, as is the case in many environments.

Now, since we are considering a situation where equation \eqref{Equation_dilatation} reduces to
\begin{equation}
\frac{\dd \vartheta}{\dd t} = - \tr(\vec{G}^2),
\label{Equation_dilatation_simplified}
\end{equation}
in the following it will matter to have information on the statistical properties of the trace of $\vec{G}^2$. For this purpose, let us relate it to more familiar quantities, whose statistical properties have been previously studied.
As detailed in, for example, \cite{IshiharaEtAl09,Meneveau11}, in small-scale hydrodynamical turbulence, the instantaneous energy dissipation rate defined as ($\nu$ is the fluid kinematic viscosity, and summation over repeated indices is implied)
\begin{equation}
\epsilon=2 \nu \mathcal{D}_{ij} \mathcal{D}_{ij},
\end{equation}
displays large levels of intermittency, being almost log-normal. Furthermore, simulations of \cite{YeungPope89,YeungEtAl06} show that the statistics of the pseudodissipation
\begin{equation}
\epsilon'=\nu G_{ij} G_{ij},
\end{equation}
is even closer to log-normality. For this reason, $\epsilon'$ is often preferred over $\epsilon$ in Lagrangian stochastic models \citep[e.g.][]{Sawford91,Lamorgese07}.
Meanwhile, the enstrophy,
\begin{equation}
\xi=\omega_i \omega_i,
\end{equation}
tends to be more intermittent than $\epsilon$ \citep[e.g.][]{ZeffEtAl03}.
We mention these three quantities because, with the decomposition \eqref{def:G} of $\vec{G}$, we have the identities
\begin{equation}
\renewcommand\arraystretch{2.} 
\begin{array}{rl}
\displaystyle  \tr(\vec{G}^2) & \hspace{-0.25cm} \displaystyle = \frac{\epsilon'}{\nu}-\xi,\\
& \hspace{-0.25cm} \displaystyle = \frac{1}{2} \left(\frac{\epsilon}{\nu}-\xi\right),
\end{array}
\label{PhysicalInterpretationOfTrG2}
\end{equation}
and given that $\epsilon$, $\epsilon'$ and $\xi$ are intermittent, these identities suggest that $\tr(\vec{G}^2)$ is also an intermittent field.

\subsection{Vorticity field}
\label{sec:Vorticity}

Taking the curl of the equation of motion \eqref{EulerEquation} yields the evolution equation for the vorticity field
\begin{equation}
\frac{\dd \vec{\omega}}{\dd t} = \vec{G} \cdot \vec{\omega} - \vartheta \vec{\omega} + \frac{\vec{\nabla} \rho \times \vec{\nabla} p}{\rho^2} + \vec{\nabla} \times \left( \frac{\vec{j} \times \vec{B}}{\rho} \right).
\label{Equation_Omega}
\end{equation}
Then in view of the decomposition \eqref{def:G}, together with \eqref{def:Omega} the first term in the right hand-side simplifies further as
\begin{equation}
\vec{G} \cdot \vec{\omega} = \vec{\mathcal{D}} \cdot \vec{\omega}.
\label{GdotOmegaEqualsDdotOmega}
\end{equation}
This term represents vortex stretching along the vortex axis. In the incompressible case, stretching along the vortex axis is automatically accompanied by contraction perpendicular to the axis, and by conservation of angular momentum, this perpendicular contraction leads to an increase of the angular velocity. In the compressible case, a new term enters the picture: the term $- \vartheta \vec{\omega}$, which represents the effect of three dimensional contraction, i.e., contraction along the vortex axis together with contraction perpendicular to the axis. Hence, the sum of the first two terms represents the effect of contraction perpendicular to the vortex axis, which again, by conservation of angular momentum, leads to an increase of the angular velocity.
The other terms in the right-hand side correspond to the contribution from baroclinity and from the Lorentz force respectively.

We are now going to adopt the same strategy as in \cite{DurriveEtAl20}. Indeed, we are going to build our model using simplifying assumptions, but since we will ultimately add several free phenomenological parameters (regularization lengths, Hurst parameters, arbitrary integration regions, and intermittency parameters), the obtained formulae will potentially mimic turbulent (magneto-)fluids in regimes beyond the framework they were inspired from. Sceptic readers at this stage are invited to have a look at the brief review of section 3.4 in \cite{DurriveEtAl20}, which stresses that this strategy turns out to be very fruitful in the well-studied incompressible hydrodynamical case of \cite{ChevillardEtAl10}. It is these successes that brought us in \cite{DurriveEtAl20} to extend the method of \cite{ChevillardEtAl10} to the magnetized case, and now to the compressible case.
Hence as in \eqref{Equation_dilatation}, assuming that it will not be as restrictive as it seems, let us neglect the back-reaction due to the Lorentz force, and consider a barotropic fluid, i.e. a fluid for which thermal pressure, $p$, is a function of density $\rho$ only. Then the last two terms of \eqref{Equation_Omega} vanish and, using mass conservation \eqref{MassConservation}, that equation may be simply written as the Beltrami equation \citep{WuMaZhou07}
\begin{equation}
\frac{\dd}{\dd t} \left(\frac{\vec{\omega}}{\rho}\right) = \vec{\mathcal{D}} \cdot \frac{\vec{\omega}}{\rho}.
\label{Equation_OmegaOverRho}
\end{equation}
Physically, the interpretation of this relation is that the quantity $\vec{\omega}/\rho$ moves as if it were frozen into the fluid, i.e. eddies are transported with the fluid, because this equation has the same form as the equation describing the evolution of the Lagrangian separation vector between two fluid particles \citep[e.g.][]{Rincon19}.

\subsection{Magnetic field}

The induction equation in the perfectly conducting limit reads
\begin{equation}
\partial_t \vec{B} = \vec{\nabla} \times (\vec{v} \times \vec{B}).
\end{equation}
Expanding the curl of the cross product on the right-hand side, this may be expressed using the velocity gradient and the dilatation field as
\begin{equation}
\frac{\dd \vec{B}}{\dd t} = \vec{G} \cdot \vec{B} - \vartheta \vec{B}.
\label{InductionEquation}
\end{equation}
With mass conservation \eqref{MassConservation} this can then be written in a similar way as the evolution equation for the vorticity \eqref{Equation_OmegaOverRho}, namely
\begin{equation}
\frac{\dd}{\dd t} \left(\frac{\vec{B}}{\rho}\right) = \vec{G} \cdot \frac{\vec{B}}{\rho}.
\label{Equation_BOverRho}
\end{equation}
This may also be interpreted as a frozen-in condition, but for the magnetic field divided by density this time, which corresponds to Alfv\'en's theorem \citep[e.g.][]{Rincon19}.
The main difference between \eqref{Equation_OmegaOverRho} and \eqref{Equation_BOverRho} comes from the fact that $\vec{\Omega} \cdot \vec{\omega}=\vec{0}$, while generally $\vec{\Omega} \cdot \vec{B} \neq \vec{0}$, so that \eqref{GdotOmegaEqualsDdotOmega} does not apply for $\vec{B}$. Physically, a vortex cannot rotate its own axis, whereas it can rotate magnetic field lines.

\subsection{Short-time dynamics}
\label{sec:ShortTimeDynamics}

The way we wrote the equations in the previous sections is well-thought-out: The form of the evolution equations for the vorticity field and for the magnetic field, \eqref{Equation_OmegaOverRho} and \eqref{Equation_BOverRho}, most naturally generalize their incompressible counterparts, equations (26) and (39) of \cite{DurriveEtAl20}, respectively. Indeed, written like this, one simply needs to replace $\vec{\omega}$ by $\vec{\omega}/\rho$, and $\vec{B}$ by $\vec{B}/\rho$, to go from the incompressible to the compressible case. But this similarity is more than a simple formal convenience: it prompts us to reuse directly the same reasoning as the one used in \cite{DurriveEtAl20} to derive the velocity and magnetic field models. This is what we are going to do, and the first step is to solve the dynamical equations as follows.

Let us consider a short time interval $t-t_0$, where $t_0$ is some arbitrary initial time (only the time difference will matter). By `short' we mean that it is short enough for the velocity gradient $\vec{G}(t,\vec{x})$ to not vary much during this period, i.e. such that it remains roughly constant, equal to its value at time $t_0$. Now since the above equations, governing the time evolution of the density, the dilatation, the vorticity and the magnetic field, are written in Lagrangian form (with the material derivative), we may integrate them as follows.

The dilatation equation \eqref{Equation_dilatation_simplified} is straightforward to integrate since in this framework $\vec{G}$ is a mere constant. Hence
\begin{equation}
\vartheta(t,\vec{x}) =\vartheta(t_0) - (t-t_0) \tr[\vec{G}(t_0)^2].
\label{dilatation_ShortTimeEvolution}
\end{equation}
Note that in order to improve the presentation we make the $\vec{x}$ dependencies implicit in the right-hand sides of the expressions throughout this section.

Then, plugging this expression for $\vartheta$ into the mass conservation \eqref{MassConservation} yields an equation for $\rho$ which is readily integrated. From this we find that the density field increases (or decreases) locally according to
\begin{equation}
\rho(t,\vec{x}) = e^{-(t-t_0) \vartheta(t_0) + \frac{(t-t_0)^2}{2} \tr[\vec{G}(t_0)^2] } \rho(t_0).
\label{rho_ShortTimeEvolution}
\end{equation}

Similarly, we may integrate equation \eqref{Equation_OmegaOverRho} to obtain the vorticity field. Since the tensor $\mathcal{D}$ is just the symmetric part of $\vec{G}$, which we assume equal to its value at time $t_0$, we may directly set $\mathcal{D}$ to its value at $t_0$, thereby finding
\begin{equation}
\vec{\omega}(t,\vec{x}) = \frac{\rho(t)}{\rho(t_0)} e^{(t-t_0) \vec{\mathcal{D}}(t_0)} \cdot \vec{\omega}(t_0).
\label{vorticity_ShortTimeEvolution}
\end{equation}
The matrix exponential corresponds to vortex stretching, and the overall factor (the ratio of density at different times, which is absent in the incompressible case), is due to local expansion or contraction.

Likewise \eqref{Equation_BOverRho} gives the short-time evolution of the magnetic field
\begin{equation}
\vec{B}(t,\vec{x}) = \frac{\rho(t)}{\rho(t_0)} e^{(t-t_0) \vec{G}(t_0)} \cdot \vec{B}(t_0).
\label{magneticField_ShortTimeEvolution}
\end{equation}
The exponential now corresponds to both stretching and also to shear. The overall factor is again a weight due to local expansion or contraction.
Note that in this procedure we assume that stretching dominates over advection, as detailed in \cite{DurriveEtAl20}. We will come back to this point in section \ref{sec:FurtherImprovements}.

\section{A model for compressible hydrodynamical turbulence}
\label{sec:Hydro}

In this section we focus on hydrodynamics. Magnetic fields will be discussed afterwards, in section~\ref{sec:MHD}.

At the heart of our model is a procedure by which we transform expressions such as those derived in section~\ref{sec:Dynamics} into random fields `adapted to mimic turbulence'. Let us illustrate this vague phrase by considering the Helmholtz decomposition~\eqref{def:v} as an example. As such, this expression does not exhibit the features we expect from a turbulent field: where is the injection scale, the dissipative scale, or even where is the randomness? Obviously, these would emerge explicitly, somehow, if we were able to completely solve the Navier-Stokes equations, since the latter embody turbulence. But as long as we are unable to do so, we only have the option of mimicking turbulence. Hence, let us try to tweak a little bit our formulae, notably the Helmholtz decomposition, in order to introduce, be it in an ad hoc manner, these important and useful features.

This is the very purpose of the procedure we are about to detail: We suggest five slight changes to bring to the dynamical equations from the previous sections, in order to introduce phenomenologically features relevant to stochastic representations of some turbulent states. Now because this procedure consists of several steps, we suggest calling it for short `turbulization', since it will make our formulae `look more turbulent'.

First, in section~\ref{sec:turbulization}, we detail what this turbulization procedure consists of, by focusing on a specific example. Then, in section~\ref{sec:InitialState}, we explicit the initial state that we think is most relevant for the astrophysical context, notably of the interstellar medium. Finally, in section~\ref{sec:DeformedState}, we show what the full deformed state is (i.e. the result of dynamically evolving the initial state), which constitutes our model for the turbulent state.

\subsection{turbulization}
\label{sec:turbulization}

The ideas underlying turbulization originate from \cite{ChevillardEtAl10}. Here we formalize this procedure, as in \cite{DurriveEtAl20}, in the sense that we decompose it into five explicit steps. To improve the presentation, we illustrate these steps with the example of the compressible component of velocity, $\vec{v}_c$, which includes the dilatation field, $\vartheta$. The expressions of $\vec{v}_c$ and $\vartheta$ are given by \eqref{def:vc} and \eqref{dilatation_ShortTimeEvolution}, respectively (because the short-time expressions are what matters from now on), which we reproduce here for convenience
\begin{equation}
\renewcommand\arraystretch{2.} 
\begin{array}{l}
\displaystyle \vec{v}_c(\vec{x}) = \frac{1}{4 \pi} \int_{\mathbb{R}^3} \frac{\vartheta \vec{r}}{r^3} \dd V,\\
\displaystyle \vartheta(t,\vec{x}) =\vartheta(t_0) - (t-t_0) \tr[\vec{G}(t_0)^2].
\end{array}
\label{Dilatation_And_Vc}
\end{equation}
Let us make the five following modifications to the expressions in~\eqref{Dilatation_And_Vc}:

(i) \textit{Injection scale parameter:} Instead of integrating throughout the whole of three-dimensional space, we limit the integration region to the vicinity of the considered position $\vec{x}$, i.e. we make the substitution
\begin{equation}
\mathbb{R}^3 \rightarrow \mathcal{R},
\end{equation}
where the region $\mathcal{R}$ enables us to control the typical size of the region of correlation. Indeed, in order to compute the velocity $\vec{v}_c$ at some position $\vec{x}$ (for simplicity, or if the turbulence is isotropic, say $\mathcal{R}$ is a ball of radius $L$ centered at this position $\vec{x}$), only the points inside $\mathcal{R}$ will matter because points further away than distance $L$, namely such that $r \equiv |\vec{x}-\vec{y}|>L$, do not enter in the integral (recall that the variable of integration is $\vec{y}$). Of course this spatial decorrelation happens smoothly: the ball used to compute $\vec{v}_c$ at a position $\vec{x}'$ close to $\vec{x}$, namely such that $|\vec{x}'-\vec{x}| < L$, will overlap with the one used to compute $\vec{v}_c$ at $\vec{x}$, but the volume of overlap decreases as the separation $|\vec{x}'-\vec{x}|$ increases. Hence, this large-scale cut-off parameter $L$ corresponds to some correlation length, and acts, in the turbulence terminology, as the injection scale, controlling the typical size of the largest eddies.

(ii) \textit{Roughness/wildness parameter:} The expression of $\vec{v}_c$ in \eqref{Dilatation_And_Vc} is the convolution of $\vartheta$ and a power-law kernel. The power of this power law is equal to~$-3$, which stems from the fact that we are considering a three dimensional space. However, fractal geometry demonstrates that considering non-integer dimensions is a powerful tool to describe a wider range of phenomena, in particular non-smooth shapes such as those typically appearing in turbulence. This approach to modeling originates from \cite{MandelbrotVanNess68,Mandelbrot72}. In practice, for our purpose, we will make use of this tool in the simplest manner: let us simply generalize the power law of the kernel in $\vec{v}_c$ to an (a priori) arbitrary value $h$, i.e. we make the substitution
\begin{equation}
r^{-3} \rightarrow r^{-2 h}.
\label{substitution_regularization}
\end{equation}
As is customary in the literature related to fractals, we call this free parameter the Hurst parameter\footnote{In fact, in order to lighten our expressions, our definition of the Hurst parameter is shifted by a constant and with a sign reversed compared to the standard definition in, for example, Fractional Brownian Motion.}, hence the choice of the letter `h'.
Physically, $h$ quantifies the `wildness' of the turbulence, in Mandelbrot's wording. Geometrically, this parameter basically controls how rough the field is, how abrupt spatial variations are. Accordingly, the power spectrum of the field is very sensitive to this parameter.

(iii) \textit{Dissipation scale parameter:} As it happens, the aforementioned kernel is singular, meaning that it diverges as $r$ goes to $0$. In order to avoid difficulties due to divergences and to ensure differentiability of the fields we are about to build, we eliminate this singularity by introducing a regularization parameter $\eta$, which simply means that we make the substitution
\begin{equation}
r \rightarrow (r^2+\eta^2)^{1/2},
\end{equation}
in the denominator of the kernel. Physically, the important advantage of introducing this small-scale cut-off is that we now gain a degree of freedom corresponding to dissipation. For this reason, we use the notation\footnote{Compared to \cite{DurriveEtAl20}, we now prefer the notation $\eta$ instead of $\epsilon$, to keep the letter $\epsilon$ for dissipation rates.} $\eta$ by analogy with the Kolmogorov lengthscale, $\eta_K$, below which viscosity dominates and the turbulent kinetic energy is dissipated into heat. This scale is usually defined as
\begin{equation}
\eta_K=\left(\frac{\nu^3}{\langle \epsilon \rangle}\right)^{\frac{1}{4}},
\end{equation}
where $\nu$ is the fluid kinematic viscosity and $\langle \epsilon \rangle$ is the mean rate of dissipation of kinetic energy per unit mass. However, beware that our parameter $\eta$ is inspired from $\eta_K$, but it is not necessarily equal to it. We use it as a free parameter, which broadly takes into account, in a phenomenological manner, all the various dissipative effects not accounted for otherwise.

(iv) \textit{Randomization and large-scale features:} We choose `initial' conditions, i.e. prescriptions for the fields $\rho(t_0),\vartheta(t_0),\vec{\omega}(t_0)$ and $\vec{B}(t_0)$. Later in the text, we will also introduce the initial current density, $\vec{j}(t_0)$, and we also need to prescribe a value for $\vec{G}(t_0)$, but this will be deduced from the aforementioned fields. 
This fourth step brings a lot of freedom in the modeling. It may give rise to a wide class of models, depending on the choices made at this level. This versatility is an asset of our approach. We may distinguish two radically different types of choices for the initial state of each field: we may either set it to a random field, in order to introduce fluctuations in the model, or we may choose it to be a smooth field, in order to inject some large-scale features or boundary effects (for example a density profile with a sharp transition between the core and the outskirt of a molecular cloud).
We will detail all our choices for the initial conditions of the present model in section \ref{sec:InitialState}. But to finish illustrating turbulization with the example \eqref{Dilatation_And_Vc}, let us mention that in this case this fourth step consists in the substitutions
\begin{equation}
\renewcommand\arraystretch{2.} 
\begin{array}{l}
\displaystyle \vartheta(t_0) \rightarrow \widetilde{\vartheta}_g,\\
\displaystyle \vec{G}(t_0) \rightarrow \widetilde{\vec{G}}_g.
\end{array}
\end{equation}
We will explicit what $\widetilde{\vartheta}_g$ and $\widetilde{\vec{G}}_g$ stand for in section \ref{sec:InitialState}, and for now we simply point out that the tildes indicate that they are random fields, with Gaussian statistics as indicated by the subscript $g$.

(v) \textit{Intermittency parameter:} The expression of $\vec{v}_c$ in \eqref{Dilatation_And_Vc} is kinematic: it describes how the velocity field varies under a local expansion or contraction, independently of the origin of this expansion or contraction. It is a geometrical relation. On the contrary, the expression of $\vartheta$ in \eqref{Dilatation_And_Vc} is dynamical\footnote{Actually in this particular example we consider a very simple version of the full equation \eqref{Equation_dilatation}, keeping only the quadratic self-amplification of velocity gradients, which is kinematic in nature. But this does not change the fact that the second equation in \eqref{Dilatation_And_Vc} is the integrated version of the dynamical equation \eqref{Equation_dilatation}.}, because it corresponds to the time evolution of $\vartheta$ deduced from the equation of motion \eqref{EulerEquation}. By construction (cf. section~\ref{sec:ShortTimeDynamics}), this relation holds a priori only during a time lapse of the order of the velocity gradient correlation timescale. However, assuming that after such a period of time the state of the fluid is somehow `reset' or `blurred' (by an external forcing or by some complicated non-linear process not modeled here) to a Gaussian state (cf. item~(iv) and section \ref{sec:InitialState}), in the following time step, the fluid particles will once more be deformed according to the equations displayed in section \ref{sec:ShortTimeDynamics}, during another velocity gradient correlation timescale. With this mental picture in mind, we build a model for a stationary turbulent field from the above dynamical relation, that is to say we replace the time interval $t-t_0$ by a free parameter $\tau$ \citep[this substitution is naturally called `stationarization'][]{ChevillardHDR,DurriveEtAl20}
\begin{equation}
t-t_0 \rightarrow \tau.
\end{equation}
Physically, $\tau$ corresponds to the time\footnote{At least it is so in this picture, useful to construct the model, but the subtlety is that in general $t-t_0$ and $\tau$ do not have the same dimensions, because we generalized the value of the Hurst exponent in relation~\eqref{substitution_regularization}.} during which fluid particles are deformed, i.e. stretched, contracted and sheared. Since the initial state is chosen to be Gaussian (cf. item~(iv) and section~\ref{sec:InitialState}), the longer this deformation lasts, the more departure from Gaussianity we expect. Therefore, the parameter $\tau$ controls the intermittency of the fluid, i.e. it quantifies how non-Gaussian the field is, and is referred to as the intermittency parameter.

\ \\
Conclusion: turbulization consists of the five above steps, and applied to the system of equations \eqref{Dilatation_And_Vc}, it results in the random fields
\begin{equation}
\renewcommand\arraystretch{2.} 
\begin{array}{l}
\displaystyle \widetilde{\vec{v}}_{c}(\vec{x}) = \frac{1}{4 \pi} \int_{\mathcal{R}} \frac{\widetilde{\vartheta} \vec{r}}{(r^2+\eta^2)^{h}} \dd V,\\
\displaystyle \widetilde{\vartheta}(\vec{x}) = \widetilde{\vartheta}_g - \tau \tr(\widetilde{\vec{G}}_g^2).
\end{array}
\label{Turbulent_Dilatation_And_Vc}
\end{equation}
Now that we have illustrated this procedure for the example \eqref{Dilatation_And_Vc} $\rightarrow$ \eqref{Turbulent_Dilatation_And_Vc}, we are ready to apply it to all the fields mentioned in section \ref{sec:Dynamics}, in order to build our models for (magneto-)hydrodynamical turbulence. But before doing so, let us explicit and justify the initial conditions that we are going to choose, such as $\widetilde{\vartheta}_g$ and $\widetilde{\vec{G}}_g$ in \eqref{Turbulent_Dilatation_And_Vc}.

\subsection{`Initial' state (undeformed state)}
\label{sec:InitialState}

The initial state represents the state of the fluid before its deformation. Therefore we will also refer to it as the `undeformed' state. In this section, we propose an initial state that is relevant for the modeling of astrophysical environments.

Inhomogeneity often plays an important role in the turbulence of astrophysical fluids, which is why we choose the initial density field to be a smooth field, i.e. not a random field, by making the substitution
\begin{equation}
\rho(t_0,\vec{x}) \rightarrow \rho_0(\vec{x}),
\label{substitution_initial_density}
\end{equation}
where $\rho_0(\vec{x})$ is, for instance, the space-averaged density profile of an interstellar molecular cloud or filament, or that of a cluster of galaxies or a filamentary structure of the cosmic web.
The substitution \eqref{substitution_initial_density} provides a way to introduce ordered, large-scale features in the modeling.

Given the stochastic nature of turbulence, it is necessary to introduce randomness at some point as well. To do so, consider two random fields, $\widetilde{\vartheta}_g$ and $\widetilde{\vec{\omega}}_g$ (tildes indicate that they are random). We randomize the initial dilatation field through the substitution
\begin{equation}
\vartheta(t_0,\vec{x}) \rightarrow \widetilde{\vartheta}_g(\vec{x}),
\label{substitution_initial_dilatation}
\end{equation}
as well as the initial vorticity field, through
\begin{equation}
\vec{\omega}(t_0,\vec{x}) \rightarrow \widetilde{\vec{\omega}}_g(\vec{x}).
\label{substitution_initial_vorticity}
\end{equation}
Let us now expose the reasoning that leads us to choose relevant explicit expressions for these two random fields.

First of all, we decide that these random fields should have Gaussian statistics, and we indicate this property with the subscripts~$g$. We make this decision based on several previous works that show that modeling turbulent fields as dynamically deformed Gaussian fields can be remarkably fruitful. Examples of such works can be found in Appendix~\ref{Appendix:Closures}, dedicated to how the question of closure is tackled in the literature.

Then, the simplest option would be to take the initial conditions $\vartheta(t_0,\vec{x})$ and $\vec{\omega}(t_0,\vec{x})$ as Gaussian white noises. In fact, this is precisely what is done in \cite{ChevillardEtAl10,DurriveEtAl20}, at least for $\vec{\omega}(t_0)$, since there is no dilatation field in their incompressible framework. The most evident reason for this is practical: it has the advantage of facilitating analytical calculations of the statistics of the process thus defined (process that we will fully explicit in section~\ref{sec:DeformedState}), as performed for instance by \cite{PereiraEtAl16}.
In addition, this choice is especially natural since Gaussian white noises are structureless (they are `pristine', in a way), and the whole point of the short-time expansion exposed in section \ref{sec:ShortTimeDynamics} is to let the equations of the dynamics determine what deformation, giving rise to the structuring, is physically relevant.

However, interestingly, while \cite{ChevillardEtAl10} ultimately choose $\vec{\omega}(t_0)$ as a Gaussian white noise, in fact they first suggest modeling it as a fractional Gaussian field. They do not push this idea further for simplicity, briefly commenting that it is a less straightforward way. 
Here in contrast, we decide to explore this second possibility because it is actually going to serve our purpose.

Indeed, on the observational side, observations of the interstellar medium show that gas density may often be well represented by a log-normal distribution \citep[e.g.][]{LevrierEtAl18}. At the same time, on the theoretical side, as detailed in \cite{DurriveEtAl20}, a convenient manner of modeling an intermittent scalar field with a long-range correlation structure is to take the exponential of a fractional Gaussian field, an idea first introduced by \cite{Kahane85} (cf. Gaussian multiplicative chaos). Finally, with our approach the density field emerges as (within a corrective function) the exponential of $\widetilde{\vartheta}_g$ (cf. equation \eqref{rhoTilde} below). For these reasons, we are naturally led to construct $\widetilde{\vartheta}_g$ as a fractional Gaussian field \citep[cf.][]{DurriveEtAl20}, namely
\begin{equation}
\widetilde{\vartheta}_g \equiv \int_{\mathcal{R}_d} \frac{\vec{r} \cdot \widetilde{\vec{s}}_g}{(r^2+\eta_d^2)^{h_d}} \dd V.
\label{vartheta_g}
\end{equation}
This form should now be familiar to the reader: it is reminiscent of the result of turbulization \eqref{Turbulent_Dilatation_And_Vc} in the example of section \ref{sec:turbulization}. Accordingly, $\mathcal{R}_d$ denotes the region of integration, by which the injection (or integral) scale is controlled, $\eta_d$ corresponds to the dissipation scale and $h_d$ regulates how rough the field is. The added subscript `d' stands for `density' because, as previously mentioned, \eqref{vartheta_g} will serve to control the statistics of the density field through \eqref{rhoTilde} below. The field $\widetilde{\vec{s}}_g$ is a Gaussian white noise vector whose components are independent of one another, zero-averaged, and with unit variance. We will call it the `seed', because it will ultimately be the only random element in the whole model, and will thus be the very origin of the randomness.

From an observational viewpoint, the choice \eqref{vartheta_g} is convenient as it introduces parameters that may readily be fitted to astrophysical data, through \eqref{rhoTilde} below. From a theoretical viewpoint, with this construction we completely anchor this modeling into the multiplicative chaos approach, pursuing in the spirit of the incompressible case \citep{ChevillardEtAl10,DurriveEtAl20}. An additional advantage is that fractional Gaussian fields are still `simple enough' to have few structures, such that with this choice we still let the short-time dynamics structure the fields act in a natural way. Indeed, the risk when introducing by hand complex features into the model (through the initial conditions in this case) is to spoil the `Navier-Stokes driven' features of the fields we build.

It remains to choose $\widetilde{\vec{\omega}}_g$. To decide, let us think in terms of degrees of freedom (DoF). We aim at constructing a stochastic representation for a three-dimensional velocity field, therefore we have three DoF to fix. But a velocity field is fully determined by its dilatation field (scalar field, so one DoF) and vorticity field (divergence-free vector field, so two DoF). So far we have fixed one DoF, by prescribing the scalar field $\widetilde{\vartheta}_g$. But to do so we introduced a vector field, $\widetilde{\vec{s}}_g$, that is three DoF. Therefore we need to construct a divergence-free vector field $\widetilde{\vec{\omega}}_g$, given that we have at hand a Gaussian white noise vector $\widetilde{\vec{s}}_g$. Seen this way, the most natural choice is to carry on with fractional Gaussian fields, and use the following expression \citep{RobertVargas08}
\begin{equation}
\widetilde{\vec{\omega}}_g \equiv \int_{\mathcal{R}_\omega} \frac{\vec{r} \times \widetilde{\vec{s}}_g}{(r^2+\eta_\omega^2)^{h_\omega}} \dd V,
\label{omega_g}
\end{equation}
which is a divergence-free Gaussian field built with $\widetilde{\vec{s}}_g$, as we were looking for. This expression also elegantly mirrors the expression of $\widetilde{\vartheta}_g$ in \eqref{vartheta_g}, in the sense that they only differ by the fact that one is derived using a scalar product while the other uses a vector product. Thus we may also infer a physical interpretation for $\mathcal{R}_\omega,\eta_\omega$ and $h_\omega$ by analogy, where now the subscript `$\omega$' means that these are vorticity-related quantities. Finally, expressions \eqref{vartheta_g} and \eqref{omega_g} also have the virtue of being differentiable, without needing to invoke the theory of distributions, contrary to Gaussian white noises.

\ \\
Having now set the initial dilatation field $\widetilde{\vartheta}_g$ and the initial vorticity field $\widetilde{\vec{\omega}}_g$, we have in fact also set the initial velocity field, according to the kinematic relation \eqref{def:v}, together with \eqref{def:vc} and \eqref{def:vr}. Now, to be fully consistent, we also need to apply turbulization to this relation. Also, since it will thus be constructed through a linear operation over the Gaussian fields $\widetilde{\vartheta}_g$ and $\widetilde{\vec{\omega}}_g$, it will also be a Gaussian random field. Let us therefore denote the initial velocity field as $\widetilde{\vec{v}}_g$. Its explicit form is given by
\begin{equation}
\widetilde{\vec{v}}_g = \widetilde{\vec{v}}_{c,g}+\widetilde{\vec{v}}_{r,g},
\label{Initial_V}
\end{equation}
with the compressional component
\begin{equation}
\widetilde{\vec{v}}_{c,g} = \frac{1}{4 \pi} \int_{\mathcal{R}_v} \frac{\widetilde{\vartheta}_g \vec{r}}{(r^2+\eta_v^2)^{h_v}} \dd V,
\label{Initial_Vc}
\end{equation}
and the rotational component
\begin{equation}
\widetilde{\vec{v}}_{r,g} = \frac{1}{4 \pi} \int_{\mathcal{R}_v} \frac{\widetilde{\vec{\omega}}_g \times \vec{r}}{(r^2+\eta_v^2)^{h_v}} \dd V.
\label{Initial_Vr}
\end{equation}
turbulization introduced $\mathcal{R}_v,\eta_v$ and $h_v$ which have analogous physical interpretation as the other ($\mathcal{R}_d,\eta_d,h_d$) parameters, but related to velocity now, as indicated by the subscript `v'.

Finally, having now set the initial velocity field, we have also set the initial velocity gradient. Indeed, applying turbulization to \eqref{def:G} yields an expression for the randomized velocity gradient, namely
\begin{equation}
\widetilde{\vec{G}}_g \equiv \widetilde{\vec{\mathcal{D}}}_g+\frac{1}{2} \widetilde{\vec{\omega}}_g \times,
\label{def:Gtilde}
\end{equation}
where, following \eqref{def:D},
\begin{equation}
\widetilde{\vec{\mathcal{D}}}_g = \widetilde{\vec{\mathcal{D}}}_{c,g} + \widetilde{\vec{\mathcal{D}}}_{r,g},
\label{def:D_Tilde}
\end{equation}
with the compressional component
\begin{equation}
\widetilde{\vec{\mathcal{D}}}_{c,g} = \frac{1}{4 \pi} \int_{\mathcal{R}_{\omega}} \hspace{-0.2cm} \frac{(r^2 \vec{I}-3 \vec{r} \vec{r}) \widetilde{\vartheta}_g }{(r^2+\eta_{\omega}^2)^{h_{\omega}}} \dd V.
\label{def:Dc_Tilde}
\end{equation}
and the rotational component
\begin{equation}
\widetilde{\vec{\mathcal{D}}}_{r,g} = -\frac{3}{8 \pi} \int_{\mathcal{R}_{\omega}} \hspace{-0.2cm} \frac{\vec{r} (\widetilde{\vec{\omega}}_g \times \vec{r}) + (\widetilde{\vec{\omega}}_g \times \vec{r}) \vec{r}}{(r^2+\eta_{\omega}^2)^{h_{\omega}}} \dd V.
\label{def:Dr_Tilde}
\end{equation}
The resulting fields $\widetilde{\vec{\mathcal{D}}}_g$ and $\widetilde{\vec{G}}_g$ are Gaussian, as indicated by the subscript $g$, because they are defined as a linear operation over a Gaussian white noise. Note that during the turbulization in $\widetilde{\vec{\mathcal{D}}}_g$ we take the same $\mathcal{R}_\omega,\eta_\omega$ and $h_\omega$ parameters as in \eqref{omega_g} because in the following $\widetilde{\vec{\mathcal{D}}}_g$ and $\widetilde{\vec{\omega}}_g$ are both going to be controlling the statistics of the vorticity field (cf. \eqref{omegaTilde} below).

\subsection{Deformed state}
\label{sec:DeformedState}

Now that our initial state is fixed, let us deform (i.e. evolve) the inital fields according to the dynamics described in section \ref{sec:ShortTimeDynamics}.

The dilatation field evolves according to \eqref{dilatation_ShortTimeEvolution}. Now applying turbulization to it, we obtain its randomized version
\begin{equation}
\widetilde{\vartheta} = \widetilde{\vartheta}_g- \tau \tr(\widetilde{\vec{G}}_{\! g}^2),
\label{varthetaTilde}
\end{equation}
where $\widetilde{\vec{G}}_{\! g}$ is evaluated using the expression \eqref{def:Gtilde}. The density field evolves according to \eqref{rho_ShortTimeEvolution}, which similarly now results in the random density field
\begin{equation}
\widetilde{\rho} = e^{-\tau \widetilde{\vartheta}_g + \frac{\tau^2}{2} \tr(\widetilde{\vec{G}}_{\! g}^2)} \rho_0.
\label{rhoTilde}
\end{equation}
In essence, this expression simply translates mass conservation. The field $\widetilde{\rho}$ corresponds to a smooth density profile $\rho_0(\vec{x})$, introduced in \eqref{substitution_initial_density}, which is perturbed by fluctuations $\widetilde{\vartheta}_g$ and $\widetilde{\vec{G}}_{\! g}$, given by \eqref{vartheta_g} and \eqref{def:Gtilde}, respectively. The form \eqref{rhoTilde} corresponds to the scalar field built at the beginning of \cite{DurriveEtAl20}, cf. their equation (9). This latter model was purely formal, inspired from multiplicative chaos only. Here, however, we are led to consider an exponential: it naturally arises from the dynamics \eqref{rho_ShortTimeEvolution} itself, such that we can now associate to the Gaussian field entering the description the physical interpretation of a dilatation field.

To construct the field \eqref{rhoTilde}, we have replaced $\vec{G}$ by a Gaussian random variable, $\widetilde{\vec{G}}_g$. But at the same time the dynamics brought us to introduce the trace of $\vec{G}^2$. Therefore we would like to have an idea of what the statistical properties of the trace of the random variable $\widetilde{\vec{G}}_{\! g}^2$ might be. At this stage, all we can say for sure is that, given that the product of Gaussian processes is in general not a Gaussian process, we at least expect $\tr(\widetilde{\vec{G}}_{\! g}^2)$ to be non-Gaussian. This fact is encouraging because, as detailed in section \ref{sec:Dilatation}, we indeed expect the trace of $\vec{G}^2$ to be an intermittent field.

An important consequence of the fact that $\tr(\widetilde{\vec{G}}_{\! g}^2)$ is intermittent is that our density field \eqref{rhoTilde} is not exactly lognormal. It is rather the product of a lognormal distribution, due to the Gaussian part $\widetilde{\vartheta}_g$, with a non-lognormal distribution, due to $\tr(\widetilde{\vec{G}}_{\! g}^2)$. This feature may seem problematic at first sight since in section \ref{sec:InitialState} we stated that we needed to end up with a lognormal density field, based on observations. Fortunately, however, here $\tr(\widetilde{\vec{G}}_{\! g}^2)$ enters only as a corrective term, since it is a higher order term in the intermittency parameter (it is proportional to $\tau^2$). 
And in fact, it turns out that deviations from lognormality are indeed expected in high Mach number turbulence \citep{RobertsonGoldreich18}, which is an underlying assumption in our construction. Therefore, this byproduct of our construction is actually an interesting feature.

Likewise, inspired from the short-time dynamics \eqref{vorticity_ShortTimeEvolution} together with \eqref{rhoTilde}, we propose to model the turbulent vorticity field as
\begin{equation}
\widetilde{\vec{\omega}} = e^{-\tau \widetilde{\vartheta}_g + \frac{\tau^2}{2} \tr(\widetilde{\vec{G}}_{\! g}^2)}  e^{\tau \widetilde{\vec{\mathcal{D}}}_g} \cdot \widetilde{\vec{\omega}}_g.
\label{omegaTilde}
\end{equation}
The exponential of the randomized strain-rate matrix $\eqref{def:D_Tilde}$ models the random stretching of vortices (in line with the fact that turbulence may basically be defined as randomly stretched vortices, as stated in \cite{WuMaZhou07}), and the exponential containing the dilatation field models the random three-dimensional contraction of vortices (along and perpendicular to their axes). Compressibility also comes into play through $\widetilde{\vartheta}_g$ inside the matrix $\eqref{def:D_Tilde}$. 

Finally, still in the same spirit, we construct the velocity field from a `turbulent version' of the Helmholtz decomposition \eqref{def:v}, namely
\begin{equation}
\widetilde{\vec{v}} = \widetilde{\vec{v}}_{c}+\widetilde{\vec{v}}_{r},
\label{Turbulent_V}
\end{equation}
with the compressional component
\begin{equation}
\widetilde{\vec{v}}_{c} = \frac{1}{4 \pi} \int_{\mathcal{R}_v} \frac{\widetilde{\vartheta} \vec{r}}{(r^2+\eta_v^2)^{h_v}} \dd V,
\label{Turbulent_Vc}
\end{equation}
and the rotational component
\begin{equation}
\widetilde{\vec{v}}_{r} = \frac{1}{4 \pi} \int_{\mathcal{R}_v} \frac{\widetilde{\vec{\omega}} \times \vec{r}}{(r^2+\eta_v^2)^{h_v}} \dd V,
\label{Turbulent_Vr}
\end{equation}
where the same turbulization was performed, with subscripts $v$ this time. This velocity field $\widetilde{\vec{v}}$ is the deformed (i.e. evolved) version of the undeformed (i.e. initial) velocity field $\widetilde{\vec{v}}_g$ given by \eqref{Initial_V}. Indeed, as a simple sanity check, notice that when $\tau=0$, $\widetilde{\vartheta}$ given by \eqref{varthetaTilde} reduces to $\widetilde{\vartheta}_g$, and $\widetilde{\vec{\omega}}$ given by \eqref{omegaTilde} reduces to $\widetilde{\vec{\omega}}_g$ (the matrix exponential reduces to the identity matrix), and, therefore, $\widetilde{\vec{v}}$ given by \eqref{Turbulent_V} reduces to $\widetilde{\vec{v}}_g$ given by \eqref{Initial_V}.

As a final remark let us stress that in fact, for pedagogical purposes, throughout the above text we misnamed two quantities. Because we tweaked a little the exact formulae from section \ref{sec:Dynamics}, strictly speaking $\widetilde{\vartheta}$ is not the dilatation field corresponding to $\widetilde{\vec{v}}$ in \eqref{Turbulent_V} and $\widetilde{\vec{\omega}}$ is not exactly its vorticity \citep[a misnomer that was already mentioned in][]{DurriveEtAl20}. Indeed, to be more precise, $\widetilde{\vartheta}$ and $\widetilde{\vec{\omega}}$ are only fields that enter the modified Helmholtz decomposition~\eqref{Turbulent_V} at the place where the true dilatation and vorticity enter the usual (i.e. unmodified) Helmholtz decomposition~\eqref{def:v}.
Hence $\widetilde{\vartheta}$ and $\widetilde{\vec{\omega}}$ should rather be called something like pseudo-dilatation and pseudo-vorticity. They only serve as intermediate tools that enable us to model the turbulent velocity field $\widetilde{\vec{v}}$ through \eqref{Turbulent_V}. The true dilatation field and vorticity field are then given by $\vec{\nabla} \cdot \widetilde{\vec{v}}$ and $\vec{\nabla} \times \widetilde{\vec{v}}$, respectively, which can be calculated afterwards.

\subsection{Advection}
\label{sec:FurtherImprovements}

The above modeling has an important shortcoming, which was already present, but not tackled, in the incompressible model of \cite{ChevillardEtAl10}: advection has been disregarded. Indeed, the relations derived in section~\ref{sec:ShortTimeDynamics} hold as long as stretching dominates over advection, as detailed in \cite{ChevillardHDR} and in section 3.2 of \cite{DurriveEtAl20}. This issue arises because our approach is inspired from works (cf. appendix~\ref{Appendix:Closures}) developed in the Lagrangian framework, i.e. following the trajectory of fluid particles, while we work using Eulerian coordinates, i.e. with a fixed frame.
Focusing on a regime in which advection is subdominant over stretching is convenient because solving a full advection equation with a variable velocity is not easy. Fortunately, since our approach has the advantage of being intuitive, we may rely on physical arguments to make an educated guess to bypass this limitation at least partially, as follows. 

Since during the dynamics of section~\ref{sec:ShortTimeDynamics} the fluid particles are advected (while being stretched) during a short period of time, let us assume that advection occurs as if the velocity does not vary during this brief time lapse. In other words, we approximate this advection as a simple translation in the direction of the initial velocity field.
This is somewhat similar to the Zel'dovich (or ballistic) approximation in cosmology, where the trajectory of particles is approximated as a straight line in the direction of their initial velocity. Of course this analogy has its limitations since for dark matter the dynamics is that of a collisionless gas under gravity, but the fact that this approximation gives rise to the caustics describing the cosmic web is a hint that in our case taking advection into account may help make the shape of the structures in the fields we build more realistic, inducing shock-like structures.
Hence, the most natural and convenient way to take this motion into account is to advect the dilatation and vorticity fields by the `initial' velocity~\eqref{Initial_V}, i.e. we propose the expression
\begin{equation}
\widetilde{\vec{v}}(\vec{x}) = \frac{1}{4 \pi} \int_{\mathcal{R}_v} \frac{\widetilde{\vartheta}(\vec{y}-\tau \widetilde{\vec{v}}_g) \vec{r} + \widetilde{\vec{\omega}}(\vec{y}-\tau \widetilde{\vec{v}}_g) \times \vec{r}}{(r^2+\eta_v^2)^{h_v}} \dd V,
\label{eq:v_advected}
\end{equation}
as an improved version of \eqref{Turbulent_V}.

\section{Three models for compressible MHD turbulence}
\label{sec:MHD}

Section \ref{sec:Hydro} was dedicated to building a turbulent velocity field. This approach may be extended to describe magnetized turbulence. Under the assumption of incompressibility, this is what \cite{DurriveEtAl20} did. They proposed three magnetic field models, which we are now ready to generalize taking compressibility into account.

\ \\
\textit{Model 1.} In many astrophysical environments, density inhomogeneity plays an important role, which is why we introduced the smooth, large-scale density profile $\rho_0(\vec{x})$ in section \ref{sec:Hydro}. Similarly, in general turbulent magnetic fields do not only consist of homogeneous fluctuations, but they also comprise an ordered component. In view of modeling astrophysical fluids, we therefore choose the initial magnetic field entering the time evolution equation \eqref{magneticField_ShortTimeEvolution} as a smooth, ordered, field. This means that we make the substitution
\begin{equation}
\vec{B}(t_0,\vec{x}) \rightarrow \vec{B}_0(\vec{x}),
\end{equation}
where the ordered magnetic field $\vec{B}_0$ provides us with large-scale features in the modeling: it can for instance be magnetic spiral arms when modeling discs of galaxies, the X-shaped magnetic field of a galactic halo, the magnetic field compressed in the shell of the Local Bubble, or a unidirectional field as the one over the full central molecular zone of the Milky Way (cf. \cite{Planck16_IntermediateResults_LargeScaleB}, \cite{FerriereTerral14}, \cite{AlvesEtAl18}, and \cite{MangilliEtAl19} respectively).

Then, applying turbulization to the expression of the magnetic field \eqref{magneticField_ShortTimeEvolution}, we construct a first model of turbulent magnetic field, namely
\begin{equation}
\widetilde{\vec{B}}^{(1)} = e^{-\tau \widetilde{\vartheta}_g + \frac{\tau^2}{2} \tr(\widetilde{\vec{G}}_{\! g}^2)}  e^{\tau \widetilde{\vec{G}}_g} \vec{B}_0,
\label{Bmodel1}
\end{equation}
where we have used the expression \eqref{rhoTilde} for the random density field. Expression \eqref{Bmodel1} generalizes the incompressible model 1 of \cite{DurriveEtAl20} in two ways: firstly the dilatation field enters explicitly the definition of $\widetilde{\vec{G}}_{\! g}$, and secondly there is now an exponential prefactor due to compressibility, which is identical to the one entering the random vorticity field \eqref{omegaTilde}, and which we may interprete similarly as the random three-dimensional contraction of flux tubes along and perpendicular to their axes.

\ \\
\textit{Models 2 and 3.} \cite{DurriveEtAl20} introduced another idea, complementary to model 1, in order to model the magnetic field: let us apply the same procedure as above, but starting back from the magnetic Biot-Savart law \eqref{BiotSavart_magneticField}. Of course, since magnetic fields are divergence-free, there is no need to include any dilatation field now. However, this approach introduces the current density field $\vec{j}$, for which we need to find a relevant `randomized' expression $\widetilde{\vec{j}}$, analogous to $\widetilde{\vec{\omega}}$.

The current density being formally similar to the vorticity of the hydrodynamical case, we may expect a priori that a relevant expression for $\widetilde{\vec{j}}$ will be something similar to \eqref{omegaTilde}. Of course, the most appropriate way to pursue would be to derive the full dynamical equation for $\vec{j}$, i.e. the generalized Ohm's law, in order to obtain an expression for the matrix that deforms $\vec{j}$, in analogy with all that we have done so far. However, we leave this task for future work. Instead, here, we generalize models 2 and 3 of \cite{DurriveEtAl20} by making the same simple reasoning as in their section 4, i.e. we only propose a way to add compressible effects in their framework, without trying to improve their modeling otherwise. Hence, in the line of \cite{DurriveEtAl20}, we assume that the matrix field responsible for deforming $\vec{j}$ is the same as that deforming $\vec{B}$, namely $\vec{G}$. This assumption is inspired from force-free dynamics in which $\vec{j}$ is proportional to $\vec{B}$. Now, to convince ourselves that this is not as restrictive as it may seem and that the resulting model may potentially be useful beyond this force-free framework, we refer the reader either to \cite{DurriveEtAl20} or to our arguments in sections \ref{sec:Dilatation} and \ref{sec:Vorticity} about neglecting the magnetic feedback due to the Lorentz force in \eqref{Equation_dilatation} and \eqref{Equation_Omega}.

On the basis of these considerations, we propose to take inspiration from \eqref{Bmodel1} and to consider as randomized current density field the following expression
\begin{equation}
\widetilde{\vec{j}}^{(i)} \equiv e^{-\tau \widetilde{\vartheta}_g + \frac{\tau^2}{2} \tr(\widetilde{\vec{G}}_{\! g}^2)}  e^{\tau \widetilde{\vec{G}}_g} \vec{j}_0^{(i)},
\label{j_tilde}
\end{equation}
where $i$ is an index that will be useful right below to distinguish between two models, called models 2 and 3 (hence $i=2$~or~$3$). The dilatation field $\widetilde{\vartheta}_g$ is given by \eqref{vartheta_g} and the stretching/shearing matrix $\widetilde{\vec{G}}_g$ by \eqref{def:Gtilde}. By analogy with \eqref{Bmodel1}, the two exponential factors in \eqref{j_tilde} may be interpreted as random three-dimensional contraction and as random stretching and shear, respectively, and the field $\vec{j}_0^{(i)}$ corresponds to the initial state of the current density field $\widetilde{\vec{j}}^{(i)}$. We make the following choices for this initial state. As detailed in section \eqref{sec:InitialState}, two extreme cases are of particular interest: taking a random field or on the contrary an ordered field. In the first case, let us consider by analogy with $\widetilde{\vec{\omega}}$, given by \eqref{omega_g}, that the initial current density is a fractional Gaussian field. This will constitute the initial current density for our model 2, namely
\begin{equation}
\vec{j}_0^{(2)} = \int_{\mathcal{R}_j} \frac{\vec{r} \times \widetilde{\vec{s}}_g}{(r^2+\eta_j^2)^{h_j}} \dd V,
\label{j_initial_Model2}
\end{equation}
where $(\mathcal{R}_j,\eta_j,h_j)$ are the same phenomenological parameters as every time previously, but related to the current density, hence the subscript `j'. This model is well suited for small-scale turbulence since it does not contain any large-scale feature.
In the second case, let us consider an ordered, large-scale current density field, $\vec{j}_0(\vec{x})$. The initial current density for our model 3 is then
\begin{equation}
\vec{j}_0^{(3)} = \vec{j}_0,
\label{j_initial_Model3}
\end{equation}
and if need be, $\vec{j}_0(\vec{x})$ may be linked to some ordered magnetic field through Amp\`{e}re's law $\vec{j}_0 = (\vec{\nabla} \times \vec{B}_0)/\mu_0$.

Finally, applying turbulization to the magnetic Biot-Savart law \eqref{BiotSavart_magneticField}, we naturally build two models ($i=2$ or $3$) for random magnetic fields, namely
\begin{equation}
\widetilde{\vec{B}}^{(i)} = \frac{\mu_0}{4 \pi} \int_{\mathcal{R}_m} \! \! \frac{\widetilde{\vec{j}}^{(i)} \times \vec{r}}{(r^2+\eta_m^2)^{h_m}} \dd V,
\label{Bmodel2_and_3}
\end{equation}
where $\widetilde{\vec{j}}^{(i)}$ is given by \eqref{j_tilde}, and $i=2$ if the initial current density $\vec{j}_0^{(2)}$ given by \eqref{j_initial_Model2} is chosen, and $i=3$ if instead $\vec{j}_0^{(3)}$ given by \eqref{j_initial_Model3} is considered. The parameters $(\mathcal{R}_m,\eta_m,h_m)$ correspond, in direct analogy with the hydrodynamical case, to the injection scale, dissipation scale and Hurst exponent of the magnetic field.

As a final remark, let us highlight an important feature of the above modeling. The link between turbulent and ordered components differs significantly from many other turbulence models in astrophysics. Indeed, our turbulent fields are the result of the deformation of the ordered fields, i.e. the ordered fields are multiplied by some fluctuating function (or matrix), while often turbulence is modeled as a sum, namely $\vec{B}=\vec{B}_\text{ordered}+\vec{B}_\text{fluctuating}$. Our approach thus introduces a natural interaction between scales, without assuming any scale separation. Another important difference is that, as a consequence, our fluctuations are proportional to the amplitude of the ordered component. The benefits and drawbacks of these distinct modelings are worth investigating, but doing so is left for future work.

\section{Conclusion and prospects}
\label{sec:Conclusion}

In this work, we have presented an extension to the hydrodynamical turbulence model of \cite{ChevillardEtAl10} and the MHD turbulence models of \cite{DurriveEtAl20} to take into account compression. This new stochastic representation is complementary to the two previous models, as it corresponds to an opposite regime of the dynamics, namely while the former were dealing with the zero Mach number regime (incompressible, highly subsonic), our construction is best suited for the infinite Mach number regime [generalizing to low, intermediate, high or arbitrary Mach numbers is left for future work, but from the construction it is already clear that this will consist in modifying the closure relation, cf. Appendix~\ref{Appendix:Closures}, since the Mach number enters the pressure terms of \eqref{Equation_dilatation} and \eqref{Equation_Omega}]. Even though we have two more fields to consider compared to them due to compressibility, namely the dilatation and density fields, we managed to keep a very strong asset of the method: all our fields (dilatation, density, velocity, vorticity, deformation, current density, and magnetic fields) are correlated, as they should in real turbulence. Indeed, all our fields are derived from \textit{a~single} Gaussian white noise vector, $\widetilde{\vec{s}}_g$, which we call the seed because it is the element that introduces all the randomness in the model. Besides, the fields are correlated in a natural physical way, since their expressions derive from the equations of the dynamics.

\ \\
Let us mention a few prospects for this work. Several free (physically motivated) parameters emerge from the construction of our model. These parameters are interesting degrees of freedom for practical purposes, as they give more latitude to fit data, but from a theoretical viewpoint the origin of these parameters needs more investigation. We already know at this stage that they are not all independent. For instance, using a one dimensional version of his model, \cite{ChevillardHDR} showed analytically that the intermittency and Hurst parameters are linearly coupled once one imposes the 4/5 law of turbulence. We will develop this important point in a forthcoming paper, as we will explore the parameter space.

A major advantage of the present approach is that it is modular: we construct the fields gradually, adding more and more ingredients in the model, each step being physically meaningful. For simplicity in this paper we made some relatively drastic approximations during the derivation. But because the model is modular, the path to improve it is clearly marked. The construction is based on the physics (the only subtle mathematical tool is multiplicative chaos, but it is not necessary to delve into its details to pursue the modeling), such that many options are already available in the literature to have physically motivated inspiration for new closure relations to input or to change our choices regarding the initial conditions, the large-scale cut-offs and the regularizations. 
In addition, a vast literature exists (cf. references mentioned in this paper) concerning the kinematics and dynamics of the gradient of velocity. At this stage, we have not taken full advantage of the results derived there. For example we did not push further the analysis of the invariants of the velocity gradients, the so-called $(P,Q,R)$ space, which could guide us towards better choices in the modeling to improve the shape of the structures emerging in the density, velocity and magnetic fields. Nor did we analyze how the properties of our fields vary with the Reynolds number, while work such as \cite{DasGirimaji19} could for instance help to do so. Another idea would be to push one order further the expansion we made in the time variable (interpreted as the intermittency parameter) when solving for the vorticity dynamics. Extending our work to take the boundaries of the system into account is important as well, and could for example be done starting with the Biot-Savart operator of a bounded domain \citep[e.g.][]{EncisoEtAl2018}.
In that sense, improving the present model should be rather straigthforward, and each new choice will result in a new model.

Finally, let us mention that we indeed expect many further improvements in the close future, well beyond our own efforts. The mathematics community is very actively exploring the concept underlying this approach, namely (Gaussian, matrix) multiplicative chaos, notably because it has many fields of applications \citep[e.g.][for most recent work]{BrokerChiranjib20,HagerNeuman20,Lacoin20,Powell20}. More specifically, in view of building a realistic stochastic representation of turbulence, ongoing efforts include \cite{ChevillardEtAl19} who manage to build a (one-dimensional for now) stochastic field which satisfies the axioms at the core of Kolmogorov's phenomenology, as well as \cite{ReneuveChevillard20} who open the path to developing a non-stationary description.

\section*{Acknowledgments}

We warmly thank Mathieu Langer and Rodrigo Pereira for fruitful discussions. This research is supported by the Agence Nationale de la Recherche (project BxB: ANR-17-CE31-0022) and the European Research Council (Advanced Grant MIST (FP7/2017-2022, No 742719)).

\bibliographystyle{mnras}
\bibliography{BxC_theory_Part2}

\bsp

\label{lastpage}

\appendix

\section{The challenge of closure}
\label{Appendix:Closures}

The evolution of velocity gradients along Lagrangian paths \eqref{EvolutionOfG} is the central equation in our model. Physically, it bears the mechanisms deforming the fluid and giving it its non-trivial structure. In fact, as seen with a first order spatial Taylor expansion, $\vec{G}$ contains all the information on the small-scale turbulence. Mathematically, this equation is crucial since this is the place where we close our system of equations, i.e. a place where we choose what physics to put into our model.
Indeed, equation \eqref{EvolutionOfG} as such is not closed (not self-consistent): notably we need to specify the pressure term. Now, in order to give the reader a feeling for how this equation may be closed, as well as to illustrate the background material on which we based our own work, we mention in this appendix works taken from the hydrodynamics literature tackling this difficult question.

In the works we detail right below, the authors consider the following analogue of our equation \eqref{EvolutionOfG}
\begin{equation}
\frac{\dd \vec{G}}{\dd t} = -\vec{G}^2 - \vec{P} + \vec{V},
\label{EvolutionOfG_Appendix}
\end{equation}
where $\vec{P}$ is the pressure Hessian tensor
\begin{equation}
P_{ij} \equiv \frac{\partial}{\partial x_j} \left(\frac{1}{\rho} \frac{\partial p}{\partial x_i} \right),
\end{equation}
and $\vec{V}$ is a term corresponding to viscosity, which we do not need to explicit because dissipative effects are taken into account through regularizations, which is the reason why we work in the inviscid limit in section \ref{sec:Dynamics}. Note, however, that this viscous term is treated in the literature in a very similar way to what we are about to show for the pressure term.

The simplest closure consists in keeping in \eqref{EvolutionOfG_Appendix} only the term $-\vec{G}^2$ corresponding to the local self-amplification of the velocity gradient tensor along Lagrangian fluid trajectories, neglecting the non-local pressure contributions and dissipative effects. It was first introduced and studied by \cite{Vieillefosse82,Vieillefosse84,Cantwell92} and is called the Restricted Euler approximation. This approximation is theoretically valuable per se because it helps focus on the stretching mechanism, but it is also the starting point for other, more realistic, closures. The problem, however, is that it leads to the development of a singularity in finite time, i.e. a divergence of the velocity gradient. Pressure and dissipation are responsible for opposing the Restricted Euler singularity in real fluids. For instance, to eliminate this singularity, one may add a linear relaxation term as in \cite{MartinEtAl98}, though this does not work for arbitrary initial conditions.

In order to deal with the pressure Hessian term, a common approach is to split the latter as
\begin{equation}
P_{ij}=\langle P_{ij} | \vec{G} \rangle + \delta P_{ij},
\end{equation}
where $\langle P_{ij} | \vec{G} \rangle$ denotes the conditional mean of $P_{ij}$, and $\delta P_{ij}$ is a fluctuation about this mean. The former quantity is modeled through statistical approximations (for example the RFD, EGF and RDGF below), and the latter is modeled using Gaussian white noises. With this decomposition, equation \eqref{EvolutionOfG_Appendix} becomes a Langevin equation, where the fluctuating pressure term acts as a stochastic forcing term, accounting for the effects not modeled explicitly, such as perturbations from neighboring eddies, larger-scale eddies or some large-scale forcing.
As a first example of this approach, one may have a look at \cite{GirimajiPope90}, where the dynamics of the velocity gradient is modeled as a tensor-valued diffusion process, in order to specifically reproduce the log-normal statistics of pseudo-dissipation. But for the direct purpose of the present work, it is the three following examples that matter the most. 

At the origin of our model is the Recent Fluid Deformation (RFD) closure, introduced in \cite{ChevillardMeneveau06,ChevillardEtAl08,ChevillardMeneveau11}. The central idea is to approximate the conditional pressure Hessian $\langle P_{ij} | \vec{G} \rangle$ as:
\begin{enumerate}[align=left]
\item an initially isotropic tensor,
\item which is deformed by the fluid during a short period of time, a period short enough for the velocity gradient to remain constant.
\end{enumerate}
Point~(i) introduces a very simple `initial' state (i.e. before the deformation), such that the non-trivial statistical properties of $\langle P_{ij} | \vec{G} \rangle$ stem from the deformation of fluid particles by $\vec{G}$. Point~(ii) clarifies the use of the word `recent' in the name RFD.
The RFD is helpful for understanding the statistics of $\vec{G}$ \citep[e.g.][]{MoriconiEtAl14}, and to build stochastic models \citep{ChevillardEtAl10}, because it is able to prevent the Restricted Euler singularity and with it one can reproduce many well-known velocity gradient characteristics (geometric properties and anomalous scaling properties).

In parallel, another closure was suggested by \cite{WilczekMeneveau14}. These authors derived the exact analytical expression of $\langle P_{ij} | \vec{G} \rangle$ for an isotropic Gaussian velocity field and then, in order to prevent the Restricted Euler singularity due to the self-amplification term $-\vec{G}^2$, they enhanced their model by tuning their parameters with data from direct numerical simulations. Consequently, they gave this closure the name `Enhanced Gaussian Field' (EGF) closure. While this approach may seem simplistic because of the Gaussian assumption, firstly it demonstrated that a significantly anisotropic conditional pressure Hessian could result from an isotropic Gaussian velocity field (due to the inherent velocity-pressure couplings), and secondly because the resulting pressure Hessian is non-trivial, it turns out that the EGF closure accounts for a certain number of non-trivial properties of the velocity gradient, enough to compete with the RFD.

Soon afterwards, \cite{JohnsonMeneveau16} weighted the pros and cons of both approaches, and built yet another closure, the so-called Recent Deformation of Gaussian Fields (RDGF) closure, which manages to combine the advantages of both the RFD and the EGF. Now the central idea is to approximate $\langle P_{ij} | \vec{G} \rangle$ as:
\begin{enumerate}[align=left]
\item initially that of an isotropic Gaussian velocity field,
\item which is deformed by the fluid during a short period of time, a period short enough for the velocity gradient to remain constant.
\end{enumerate}
Clearly, point~(i) is inspired from the EGF (but only the exact analytical part of this model, without requiring the tuning with numerical simulations). This, compared to the RFD, improves the choice of initial condition because thus the initial conditional pressure Hessian is no longer isotropic, which relaxes the strong assumption made in the RFD. In contrast, point~(ii) is inspired from the RFD, and is the ingredient that makes the statistics of the field non-Gaussian in a most natural way, i.e. guided by the dynamics.

The above examples were taken from incompressible hydrodynamics studies. As an example of closure constructed in the context of compressible flows, let us mention the so-called Homogenized Euler model by \cite{SumanGirimaji09}. In this case, a polytropic equation of state is considered, and, most importantly, the velocity gradient $\vec{G}$ is assumed to be spatially homogeneous, as in \cite{Vieillefosse82} (hence its name referring to the Restricted Euler closure), and discard all non-local effects. The same authors later introduced an enhanced version of their model, accounting for some critical Mach-number-, Reynolds-number- and Prandtl-number-dependent physics in \cite{SumanGirimaji11}. While this model shares the same ambition as our model, namely of tackling the question of compressibility, we preferred nonetheless to take inspiration from the RDGF instead, at this stage at least.

\end{document}